\documentclass[twocolumn]{aastex6}

\usepackage{natbib}
\usepackage{graphicx}
\usepackage{amsmath}   
\usepackage{float}
\usepackage[caption=false]{subfig}
\usepackage{wrapfig}
\usepackage{gensymb}
\usepackage{url}
\usepackage{hyperref}   
\usepackage{color}
\usepackage{amssymb}
\usepackage{epstopdf}
\usepackage{calc}
\usepackage{color}

\graphicspath{{./}}
\begin{document}

\title {Is there a Disk of Satellites around the Milky Way?}

\author
{
Moupiya Maji$^{1, 2}$, Qirong Zhu$^{1,2,3}$, Federico Marinacci$^{4}$,and Yuexing Li$^{1, 2}$}	

\affil{
$^{1}$Department of Astronomy \& Astrophysics, The Pennsylvania State University, University Park, PA 16802, USA \\
$^{2}$Institute for Gravitation and the Cosmos, The Pennsylvania State University, University Park, PA 16802, USA\\
$^{3}$Harvard-Smithsonian Center for Astrophysics, Harvard University, 60 Garden Street, Cambridge, MA 02138, USA\\
$^{4}$Department of Physics, Kavli Institute for Astrophysics and Space Research, Massachusetts Institute of Technology,
Cambridge, MA 02139, USA\\
}

\email{moupiya@psu.edu} 

\begin{abstract}

 The ``Disk of satellites" (DoS) around Milky Way is a highly debated topic with conflicting interpretations of observations 
 and their theoretical models. We perform a comprehensive analysis of all dwarfs detected in the Milky Way and find that the 
 DoS structure depends strongly on the plane identification method and the sample size. In particular, we 
 demonstrate that a small sample size can artificially produce a highly anisotropic spatial distribution and a strong clustering
 of the angular momentum of the satellites. Moreover, we calculate the evolution of the 11 classical satellites with proper
 motion measurements and find that the thin DoS in which they currently reside is transient. Furthermore, we analyze two cosmological 
 simulations using the same initial conditions of a Milky Way-sized galaxy, an N-body run with dark matter only and a hydrodynamic
 one with both baryonic and dark matter, and find that the hydrodynamic simulation produces a more anisotropic distribution of 
 satellites than the N-body one. Our results suggest that an anisotropic distribution of satellites in galaxies can originate
 from baryonic processes in the hierarchical structure formation model, but the claimed highly-flattened, coherently-rotating
 DoS of the Milky Way may be biased by small-number selection effect. These findings may help resolve the 
contradictory claims of DoS in galaxies and the discrepancy among numerical simulations.
 
\end{abstract}


\maketitle

\section{Introduction}
Four decades ago, it was first reported that five bright  
satellite galaxies of the Milky Way (MW) align in a plane inclined 
to the Galactic stellar disk \citep{LyndenBell1976}, a phenomenon later dubbed as 
``disk of satellites" (DoS)  \citep{Kroupa2005} that included 11 bright MW dwarfs. Recently, it was 
claimed that 8 of these satellites co-rotate in the DoS \citep{Metz2008, 
Pawlowski2013}. Numerical simulations with the standard Lambda Cold Dark
Matter ($\Lambda$CDM) model have been largely unsuccessful to reproduce such a 
spatially-thin, kinematically-coherent structure, which has been
strongly criticized as a failure of the standard $\Lambda$CDM cosmology \citep{Kroupa2005,
Pawlowski2015methods}. 

To date, more than three dozens of dwarf galaxies have been detected around the MW 
\citep{McConnachie2012, Koposov2015},  and it was suggested that all
satellites lie in the original DoS formed by the 11 classical satellites
\citep{Pawlowski2015}. More intriguingly, it was recently reported that about
half of the satellites in Andromeda (15 out of 27) form a DoS around the host 
\citep{Pawlowski2013, McConnachie2009, Conn2013}, and that 13 out of the
15 co-planar satellites co-rotate based on line-of-sight velocities \citep{Ibata2013}.
Outside of the Local Group, one study \citep{Ibata2014} found
22 galaxies in the Sloan Digital Sky Survey (SDSS)
catalog which have diametrically opposed satellite pairs with anti-correlated velocities, and the authors suggested
that co-planar and co-rotating DoS is common in the Universe.

The origin of the DoS, however, has remained an unsolved mystery. On the one hand,
many advanced $\Lambda$CDM simulations have failed to produce such  
thin, co-rotating DoS in galaxies. While some sophisticated simulations have managed to produce an
anisotropic distribution of satellites \citep{Pawlowski2015methods, Buck2016,
Sawala2016, Zhu2016, Papastergis2016}, no consensus of coherent motion was found 
in the DoS \citep{Buck2016, Sawala2016, Bahl2014, Cautun2015a}. 

On the other hand, the interpretation of DoS from observations has been 
called into question. \cite{Buck2016} demonstrated that line-of-sight
velocities are not representative of the full 3-D velocity of a galaxy and
they cannot be used to derive coherent motion in Andromeda satellites.
Furthermore, recent investigations of the SDSS galaxies by \cite{Cautun2015b} and \cite{Phillips2015} found that 
the excess of pairs of anti-correlated galaxies is very sensitive
to sample selection and it is consistent with the random noise corresponding to an
under-sampling of the data.

In order to resolve the controversies surrounding the DoS, we reanalyze the observed 
satellites of the MW and compare them with advanced simulations. We focus on the following important issues: (1) effects of the plane 
identification method and sample size 
on the DoS properties, (2) the stability of the planar structure; and (3) effects of baryons on the distribution and evolution of satellites.  

The paper is organized as follows. In $\S 2$ we introduce the methods used in this study, including the techniques to identify the planar structure, 
the model to project future evolution of the current satellites, and the cosmological simulations with and without baryons. We present our results
in $\S 3$, namely the structural and kinematic properties
of the observed satellites using different plane identification methods and sample sizes in $\S 3.1$, the dynamical evolution of the observed 11 
classical satellites in $\S 3.2$, and the DoS structure and its evolution from two cosmological simulations in $\S 3.3$. We summarize our findings
and their implications in $\S 4$.

\section{Methods}

We use two types of techniques to analyze the present distribution of the positions of 
the observed satellites around the MW:  the Principal Component Analysis (PCA) and the Tensor of Inertia (TOI).

For our specific case of 3D positional data, PCA can be thought of as fitting an ellipsoid to the data, where the 
ratio of minor and major axis ($c/a$) indicates the anisotropy of the dwarf distribution. 
If the distribution of the dwarfs is perfectly planar, then $c/a \rightarrow 0$.
In the TOI method, which is often used in literature \citep{Allgood2006},
we calculate the moment of inertia matrix of the satellites and diagonalize it.
The eigenvalues of this matrix gives the three axes ($a,b,c$) of the fitted ellipsoid to the dwarf distribution. 
It has been argued that distant dwarfs in this distribution have a greater chance of being outliers, hence 
they should carry less weight in the TOI calculations. Here we consider three different weights for satellite
distances, namely $1, 1/r$ and $1/r^2$, respectively, as used by different groups in literature \citep{Pawlowski2015methods, 
Cautun2015b, Sawala2016}. We discuss these methods in more detail in a companion paper
(Maji et al. 2017, in prep).

Moreover, in order to investigate the stability of the DoS, we employ the galaxy dynamics software Galpy 
\footnote{http://galpy.readthedocs.io/en/latest/}
\citep{Bovy2015} to predict the future position and velocity of the observed 11 classical satellites. 
We use a realistic MW potential with three components: a power-law density profile (cut-off at 1.9 kpc) for the central bulge,
a stellar disk represented by a combination of 3 Miyamoto-Nagai potentials (MN3 model) with varying disk mass and radial
scalelength \citep{Smith2015}, and Navarro-Frenk-White \citep{Navarro1996} density profile for the dark matter halo.
We take the initial position and velocities in galactic coordinates from \cite{Pawlowski2013} and convert them
into galactocentric cartesian coordinates 
\citep{Johnson1987}.


Finally, in order to understand the origin of the DoS, we compare two cosmological simulations of a MW-size galaxy, 
one with both baryons and dark matter (hereafter referred to as ``Hydro  Simulation", \citealt{Marinacci2014}) and 
the other with dark matter only (hereafter referred to as ``DMO Simulation", \citealt{Zhu2016}). The Hydro simulation
includes a list of important baryonic physics, such as a two-phase ISM, star formation, metal cooling, and feedback from
stars and active galactic nuclei (AGN). We refer the reader to \cite{Marinacci2014} for more details on the this simulation.
The dwarf galaxies (subhalos) in the simulations are identified using the Amiga Halo Finder \citep{Knollmann2009}, a 
density-based group finder algorithm.

\section{Results}

\subsection{DoS properties with different methods and sample sizes}
\subsubsection{Structural properties}
\begin{figure}
\centering
\includegraphics[width=0.49\textwidth,clip=true,trim=0pt 0pt 0pt 0pt]{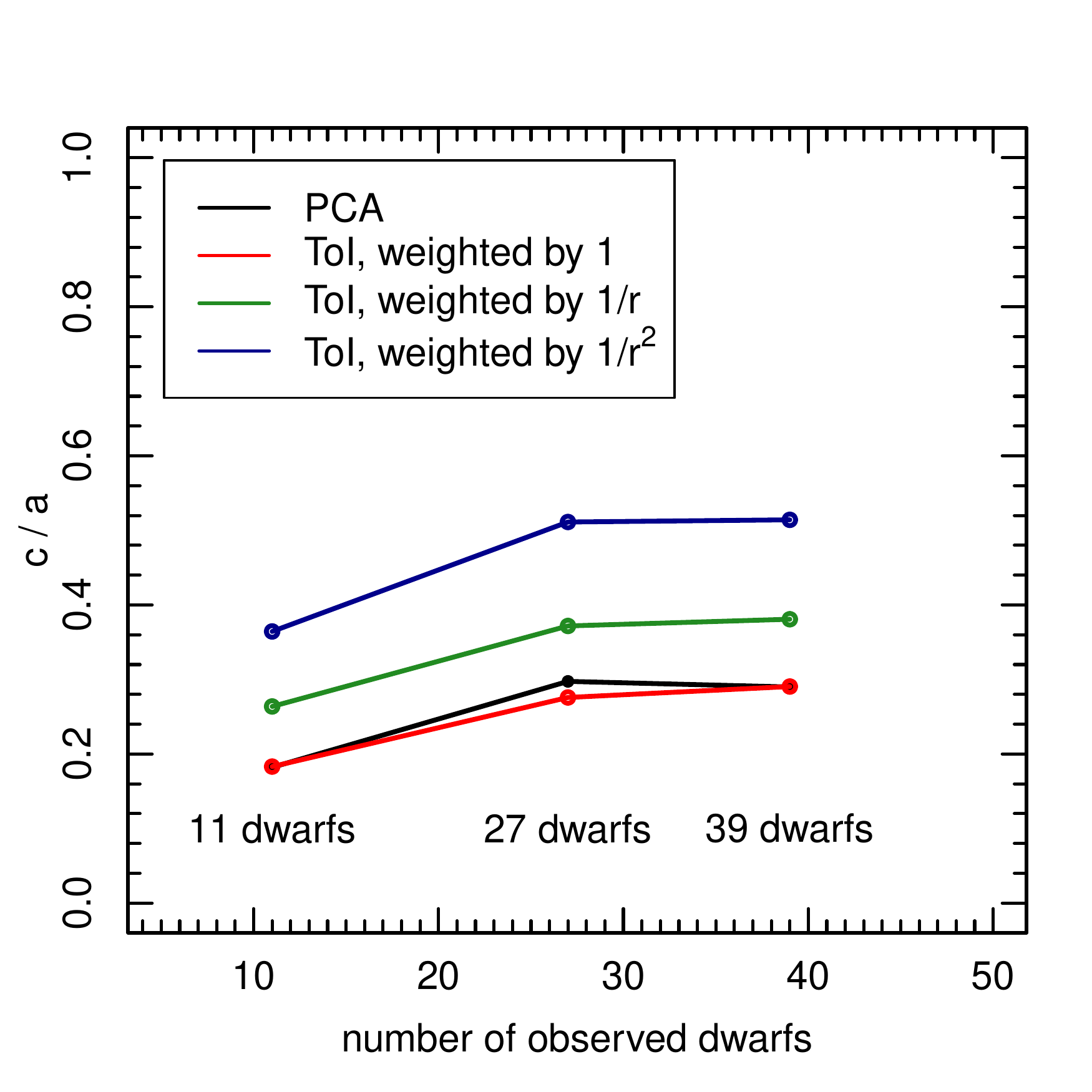}\\
\includegraphics[width=0.49\textwidth]{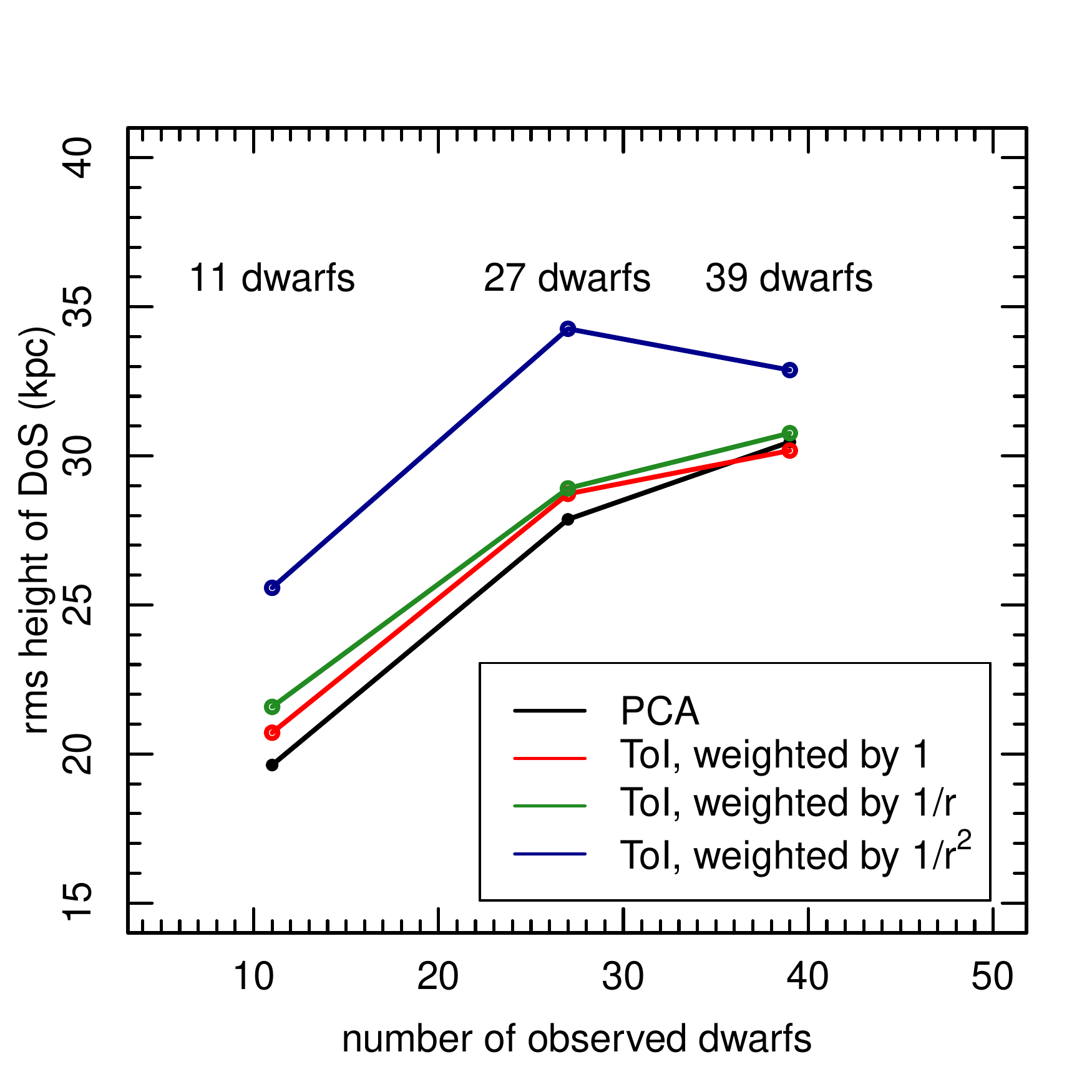}
\caption{
A comparison of the DoS structure using different sample size and plane fitting method:
``isotropy" ({\it top}) as indicated by the
ratio between semi-minor and semi-major axes, $c/a$ ($c/a=0$ means 
completely anisotropic planar distribution);
and ``thickness" 
({\it bottom}) as indicated by the root-mean-square height of the fitted plane.
The plane fitting methods include PCA and TOI with different weight function. The
complete sample includes 39
confirmed satellites of the MW \citep{McConnachie2012, Koposov2015}. 
}
\label{fig:ca_obs}
\end{figure}

A comparison of the DoS structure using different plane identification methods (PCA and TOI) 
and different sample sizes is illustrated in Figure~\ref{fig:ca_obs}. The sample of 39 currently confirmed dwarfs of 
the MW \citep{McConnachie2012, Koposov2015} includes the 11 classical satellites \citep{Kroupa2005} and the 27
most massive nearby ones in the previous analysis \citep{Pawlowski2013}. Clearly, when the satellite number 
increases from the original 11 to the full sample of 39, the ``isotropy" of
the DoS (represented by the ratio between semi-minor and semi-major axes of the principal components, $c/a$)
increases from $c/a \sim 0.2$ to $\sim 0.26$ using the PCA and unweighted TOI methods, and the 
``thickness" of the DoS (represented by the root-mean-square height of the fitted plane) increases
rapidly from $\sim 20$ kpc to $\sim 30$ kpc. For a weighted TOI with $1/r^2$ typically used in the analysis of
cosmological simulations \citep{Sawala2016},
the DoS becomes more isotropic and thicker. This figure demonstrates that the DoS structure is   
subjected 
to selection effects, which explains
the different claims reported in the literature using different methods and sample sizes 
\citep{Sawala2016, Pawlowski2015}.

\begin{figure}
 \centering
 \includegraphics[width=0.49\textwidth,,clip=true,trim=0pt 0pt 0pt 0pt]{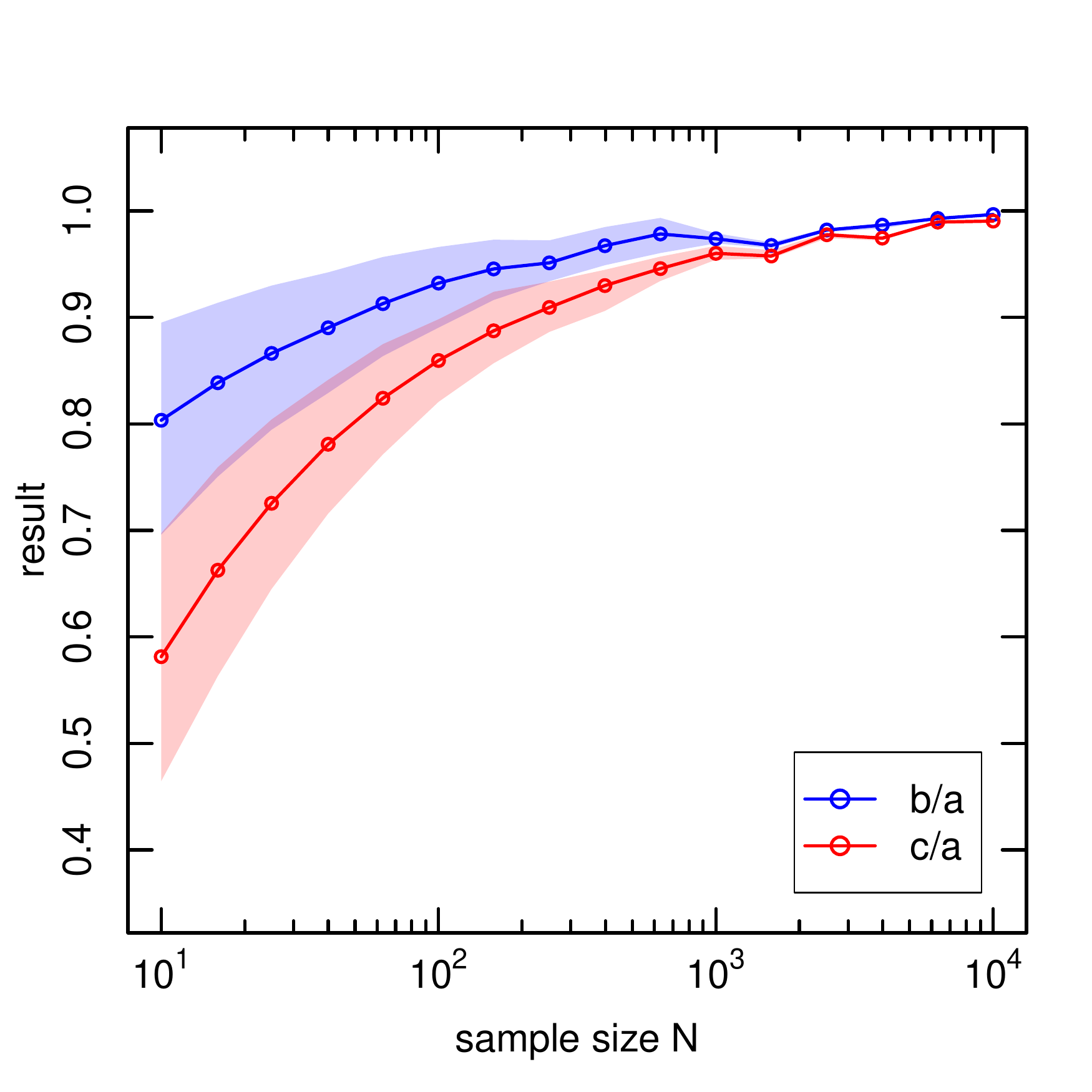}
 \caption{Effects of sample size on the anisotropy measurement of a system. The red and blue lines represent
 the $c/a$ and $b/a$ ratio of the sample, respectively, and the shaded regions indicate the $1\sigma$ error bar of the measurements.}
 \label{fig:ca_ba_number}
\end{figure}

In order to test the effect of sample size on the anisotropy measurements in more detail,
we sample an isotropic distribution (input $c/a$ and $b/a$ = 1) with $10^4$ objects.
We repeatedly draw random samples with given size
from the sphere and calculate $c/a$ and $b/a$ ratios of the sample using the unweighted TOI method.
The variation of these anisotropy ratios with the sample size is shown in Fig~\ref{fig:ca_ba_number}. For large sample size N, the 
output ratios do point to the true results of both ratios being 1. On the other hand, for small (e.g. N$\sim$10), it does not adequately sample 
the sphere, resulting in very biased estimates (for $N=10$, median $c/a = 0.58, b/a = 0.8$).
We discuss this effect in more detail in a companion paper (Maji et al. in prep.) where we place 11 satellites at their observed distances,
vary the input $c/a$ from 0.4 to 1.0 and perform a Monte Carlo simulation with $10^5$ realizations. 
We find that for all
input $c/a$ values, the output $c/a$ is consistently biased towards lower value. 
With an input $c/a \sim 0.4$, there is a 20$\%$ chance that the system has 
$c/a \lesssim 0.18$. We also find that weighted TOI method ($1/r^2$) consistently gives better result (closer to true value) compared to the
unweighted method. This analysis indicates that the system appears to be
more anisotropic when the sample size is very small because a small sample size
systematically yields a 
lower $c/a$ ratio than the true underlying anisotropy of the system.

\subsubsection{Kinematic properties}
\begin{figure*}
\centering
\includegraphics[width=0.47\textwidth,clip=true,trim=0pt 10pt 0pt 40pt]{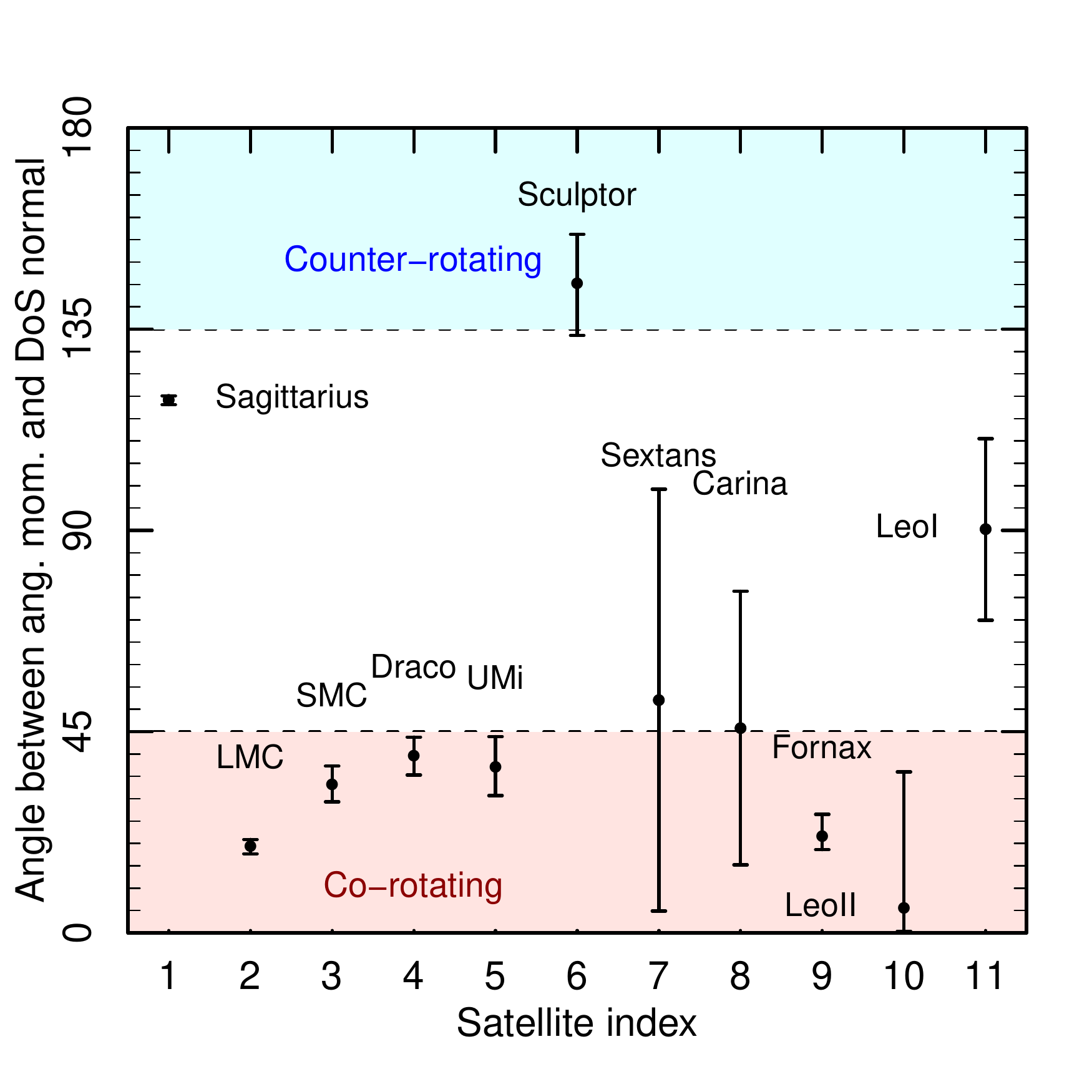}
\includegraphics[width=0.45\textwidth, clip=true,trim=0pt 0pt 0pt 60pt]{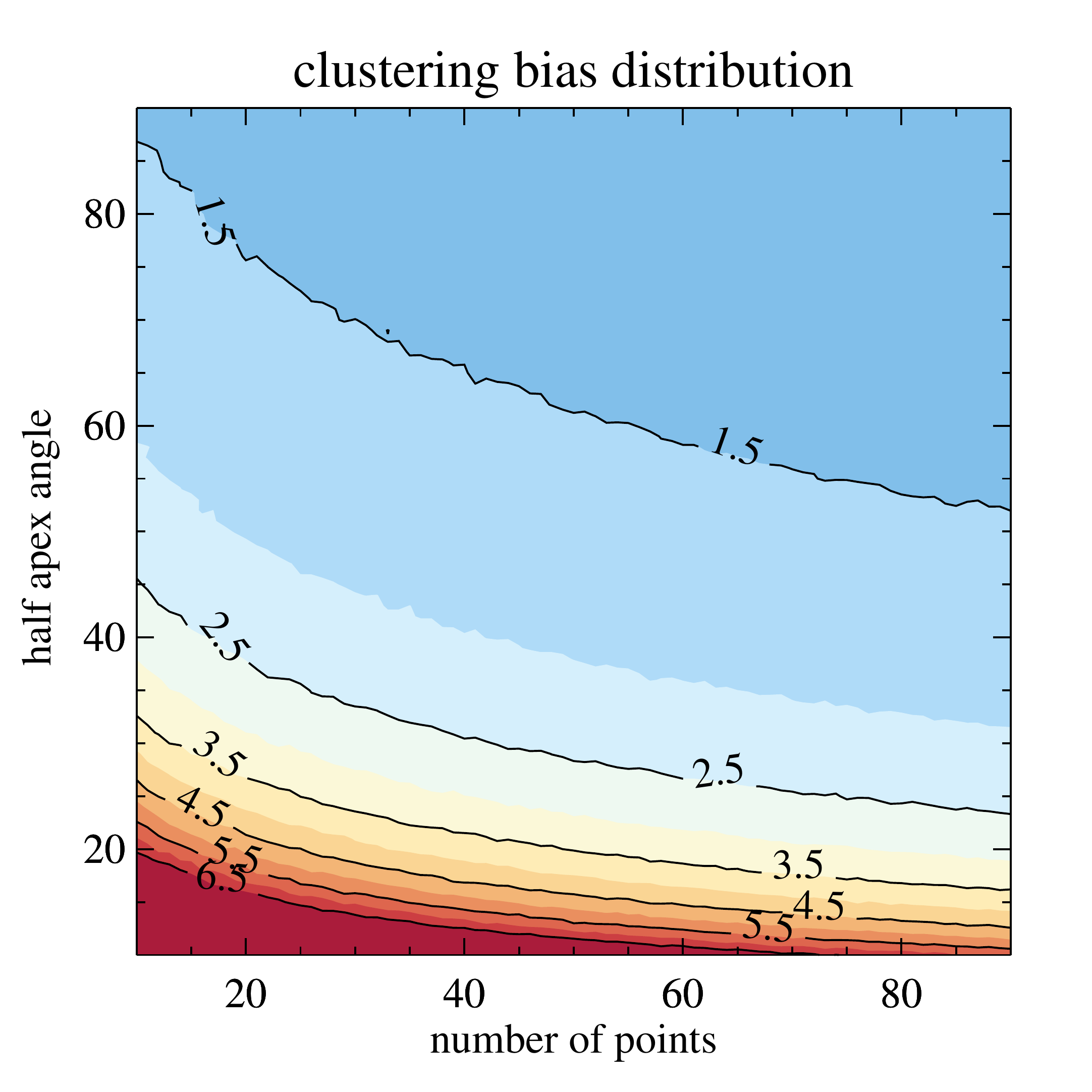}
\caption{\textit{Left panel:} 
Distribution of the angle between the satellite angular momentum and the DoS normal, with their respective error bars resulting from the
uncertainties in velocity measurements.
Satellites can be considered as corotating on the DoS if this angle is within 45 degrees (pink region) and counter-rotating if they
are within 135 - 180 degrees (green region).
\textit{Right panel:} Half apex angle of the cone vs. the number of points found in them. We draw random data points from a isotropic
point distribution in a sphere and search for clustering within different half apex angle cones in each of $10^4$ trials.
The numbers on the contours represent bias parameters. For 11 satellites in a uniform distribution, there is a 6$\%$ chance that 6
of them are clustered within 45 degree.
}
\label{fig:lndos}
\end{figure*}

In a recent study by \cite{Pawlowski2013}, it was suggested that 7 to 9 out of the 11 classical MW satellites are co-rotating because 
the angles between their angular momenta and the DoS normal of the 11 satellites fall within 45 degrees. We use the same criterion for
corotation in this study and show the angular momentum distribution of the satellites in  Figure~\ref{fig:lndos}. As shown in the left
panel of the figure 8 satellites appear in the corotation region (similar to the result of \citealt{Pawlowski2013}), but given 
the large error bars of Sextans and Carina, only 6 (LMC, SMC, Draco, UMi, Fornax and LeoII) can be robustly considered as corotating. 
However, this sample size is very small and apparent clustering can often be found in random distributions. This effect, known as the clustering
illusion \citep{clarke1946}, can lead to misinterpretation of the data as we demonstrate below.

In order to understand the significance of the co-rotation and the effect of sample size, we perform a ``clustering" test. 
Our null hypothesis is that there is no coherent motion on the DoS plane, i.e. there is no clustering of 
the angular momentum on the sphere. We use Monte Carlo simulations to numerically test the apparent clustering seen in the   
observed satellite angular momenta.  We draw $N$ random data points from a uniform distribution on a sphere and search for
clustering for each draw within a given apex angle. This experiment is repeated for $10^4$ trials.  First, we carry out this
experiment with a fixed number $N = 11$, i.e. the number of classical MW satellites.
It is found that the median number of clustered points within 45 degrees is 4, and
the chance of finding 5 or 6 clustered points within 45 degrees, similar to the clustering for observed satellites, is
$\sim 19\%$ for 5 and $\sim 6\%$ for 6 points, respectively. 
Next, we repeat the simulation with a varying number of points. To quantify the effect of sample size, we define a bias parameter as the
ratio between the observed number of clustered points and an expected number proportional to the solid angle ($S$) of the cone 
($S/4\pi\times N$). The resulting distribution of clustering from these experiments are shown in Figure 3.

This figure demonstrates that for smaller sample size, the clustering bias is significantly higher at given small angles.
A strong clustering factor (2.5 - 3.5) at $N < 20$ and angle $< 45$ can be  found due to the intrinsic 
fluctuations of random points alone. This test shows that, even though the intrinsic distribution is uniform, the
points can appear highly clustered for a small sample size. Therefore, we caution that the evidence of coherent rotation in the
11 observed satellites may not be conclusively different from that of a random data sample.

\subsection{Dynamical evolution of satellites}

\begin{figure*}
 \centering
 \includegraphics[width=0.32\textwidth]{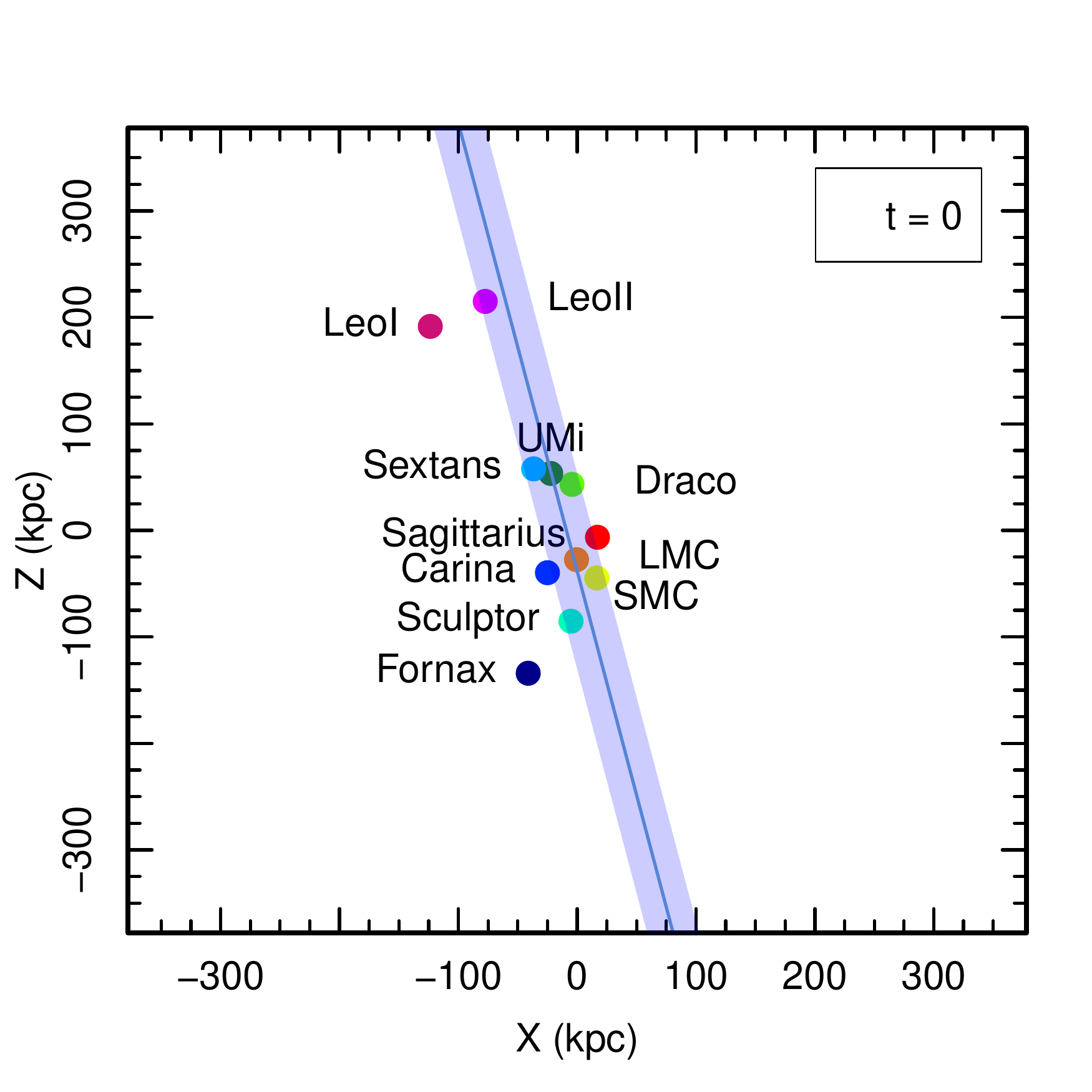}
  \includegraphics[width=0.32\textwidth]{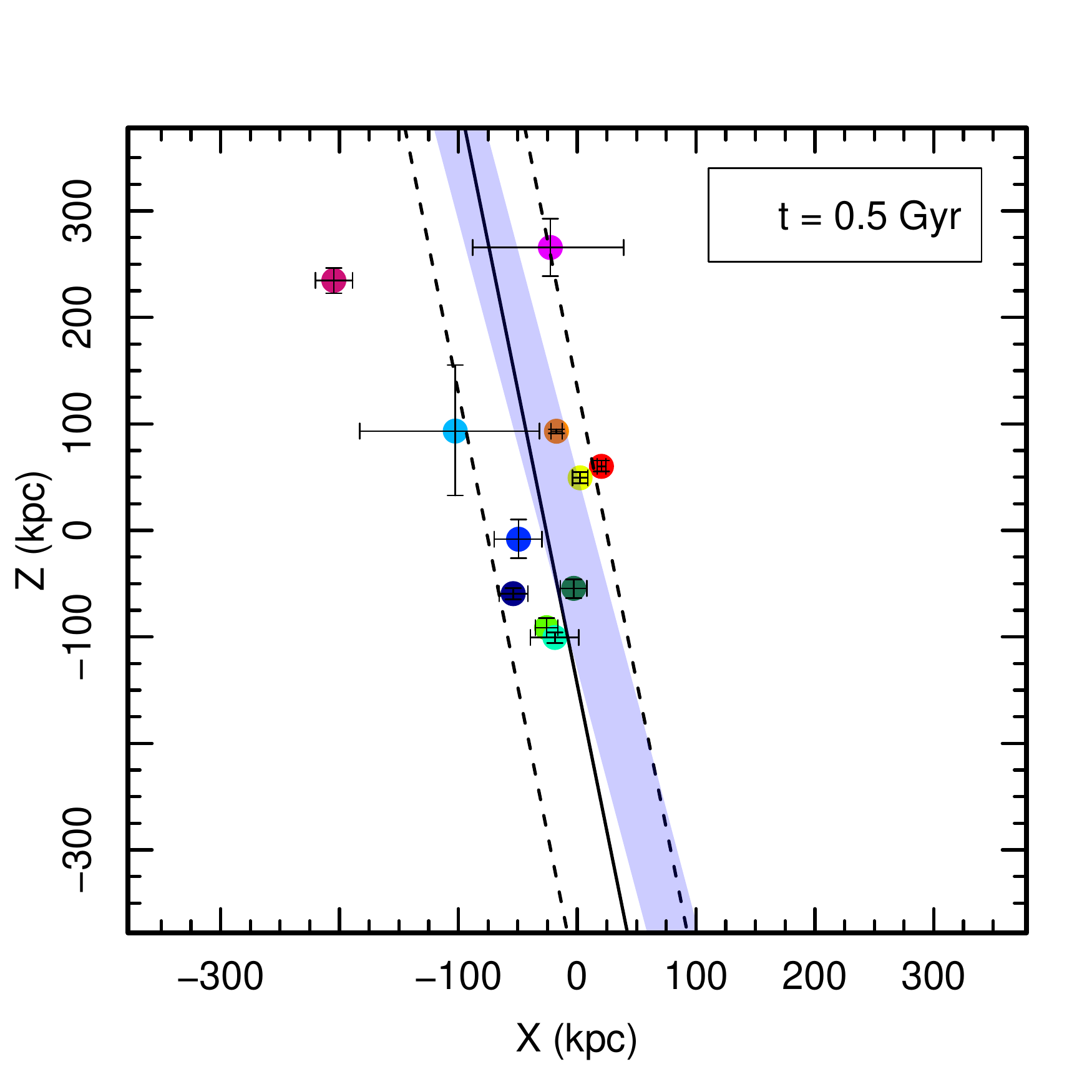}
  \includegraphics[width=0.32\textwidth]{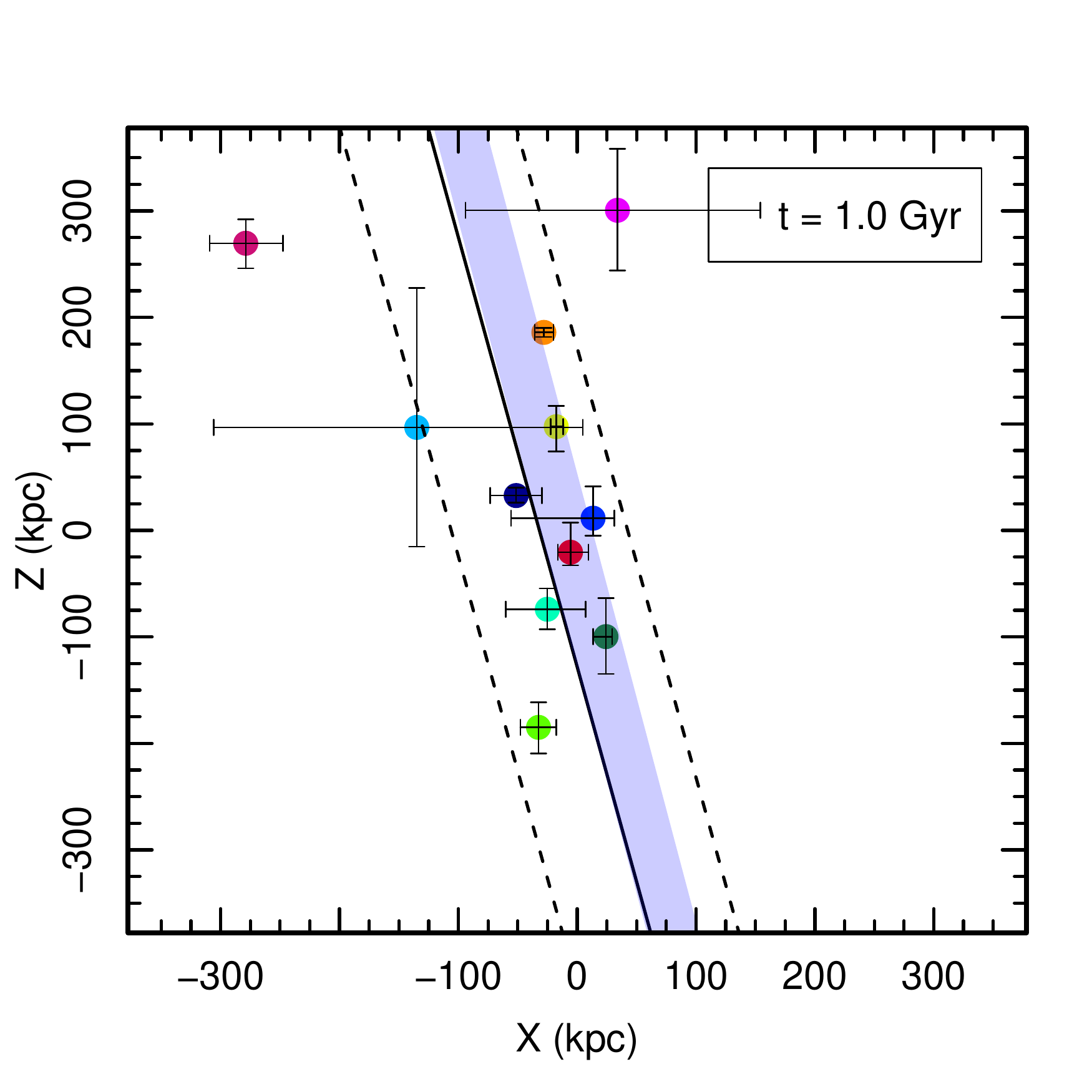}
  \caption{ Positions of 11 classical satellites in galactocentric co-ordinates at the present (left), 
  0.5 Gyr (middle) and 1 Gyr from now (right), respectively. The solid lines in each panel represent the fitted DoS at that time and the dashed
  lines represent the r.m.s. height of the plane. The blue shaded region in each panel depicts the present-day DoS.}
  \label{fig:future_mn3}
\end{figure*}

Recently,  \cite{Lipnicky2016} studied the dynamics history of the 11 ``classical" satellites and suggested that the DoS would lose its significance
in less than one Gyr in the past. In order to investigate the future evolution of DoS, we use the galactic dynamics software 
Galpy \citep{Bovy2015} to predict the future trajectories of the 11 classical satellites. 

Figure~\ref{fig:future_mn3} shows the future positions of the 11 satellites using a realistic MW potential with three components: a dark matter 
halo with the NFW density profile \citep{Navarro1996}, a central bulge with a power-law density profile cut off at 1.9 kpc, and a stellar disk 
with the MN3 potential  \citep{Smith2015}. Note that the points only represent the final positions at these times, not the detailed orbits of 
the satellites, and that nearby satellites such as Sagittarius may complete more than one orbit in 1 Gyr while distant ones such as LeoII may 
move only a fraction of their orbits. To estimate the error bars in the positions, we model the present velocities as a normal distribution 
taking as a standard deviation their present-day uncertainties. We take 1000 random samples from this velocity distribution,  calculate their
future trajectories and take the 16th and 84th percentile value (which approximate the 1$\sigma$ confidence interval) of these future position
distribution as our lower and upper error bars. Some of these satellites have large proper motion errors which propagates a significant uncertainty
in far future positions, so predictions beyond 1 Gyr are not trustable \citep{Lipnicky2016}. 
 
From this figure, we find that the 11 satellites are moving away from the present DoS at future times. The new fitted DoS is thicker 
with $c/a\sim 0.36$ (height~45 kpc) at $t=0.5$ Gyr and $c/a\sim 0.42$ (height~64 kpc) at t=1 Gyr, compared to the thin DoS 
($c/a ~0.18, \rm height~ 19.6$ kpc) at the present time. We have also explored two different MW potentials, by replacing the
stellar disk with a one component Miyamoto-Nagai potential \citep[MW2014 model]{Bovy2015}, and a NFW dark matter halo only potential,
but the resulting positions (not shown to avoid overcrowding) are very similar to those from Figure~\ref{fig:future_mn3}. These 
calculations show that, for these idealized potentials, the MW satellites tend to move away from the present DoS, increasing its
thickness and  suggesting that the current thin DoS may be a transient structure.

\subsection{Evolution of DoS isotropy in simulations}
\begin{figure*}
\centering
\includegraphics[width=0.32\textwidth]{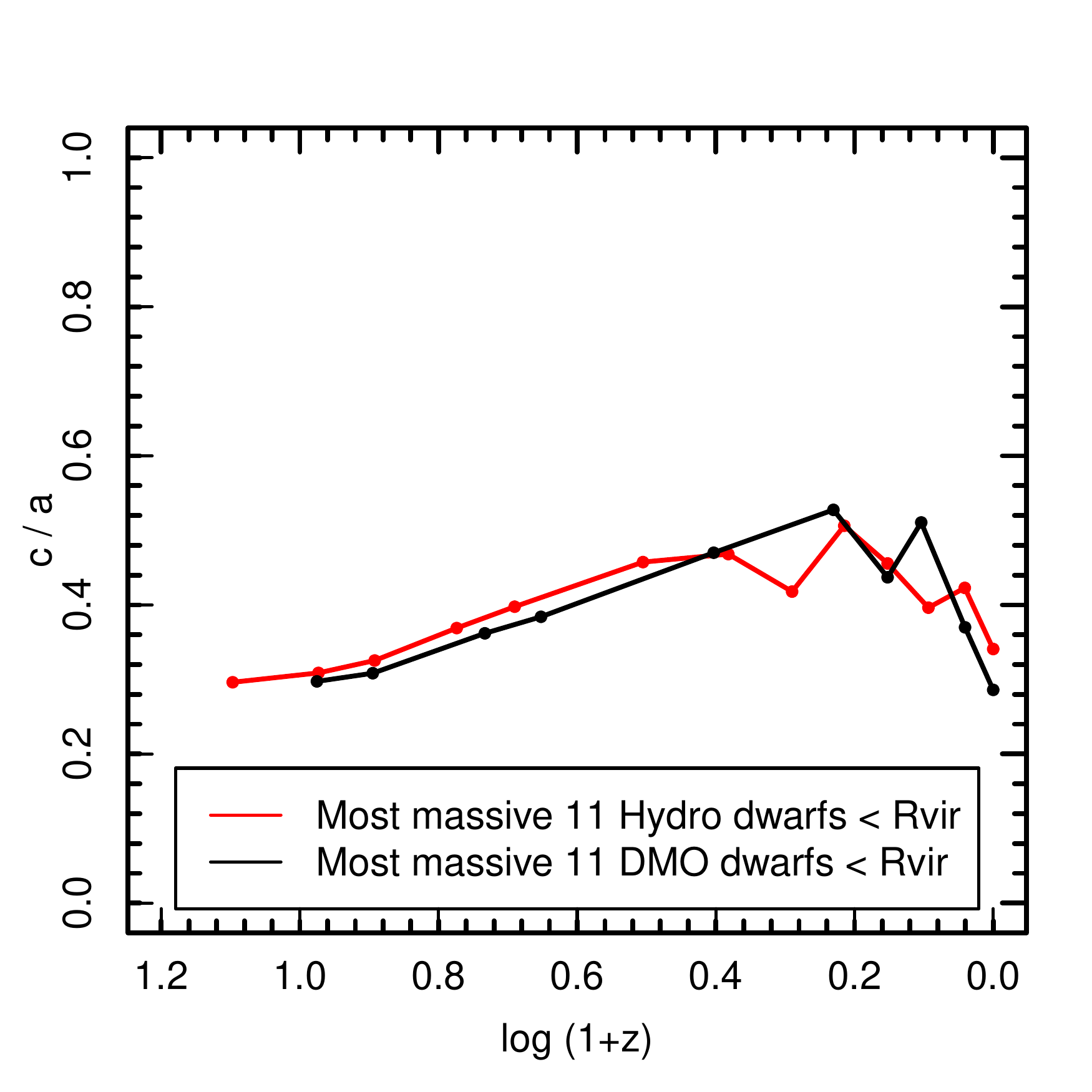}
\includegraphics[width=0.32\textwidth]{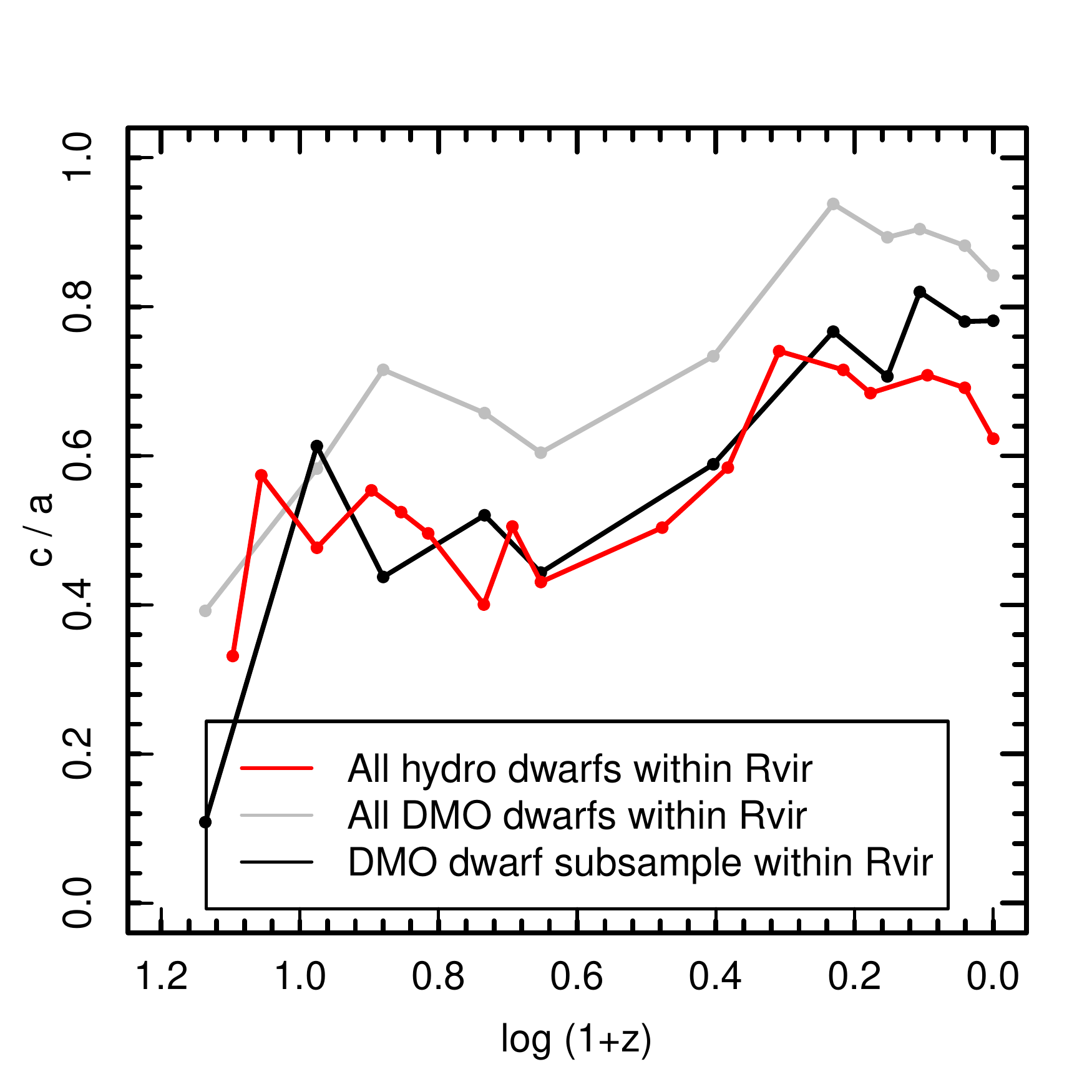}
\includegraphics[width=0.32\textwidth]{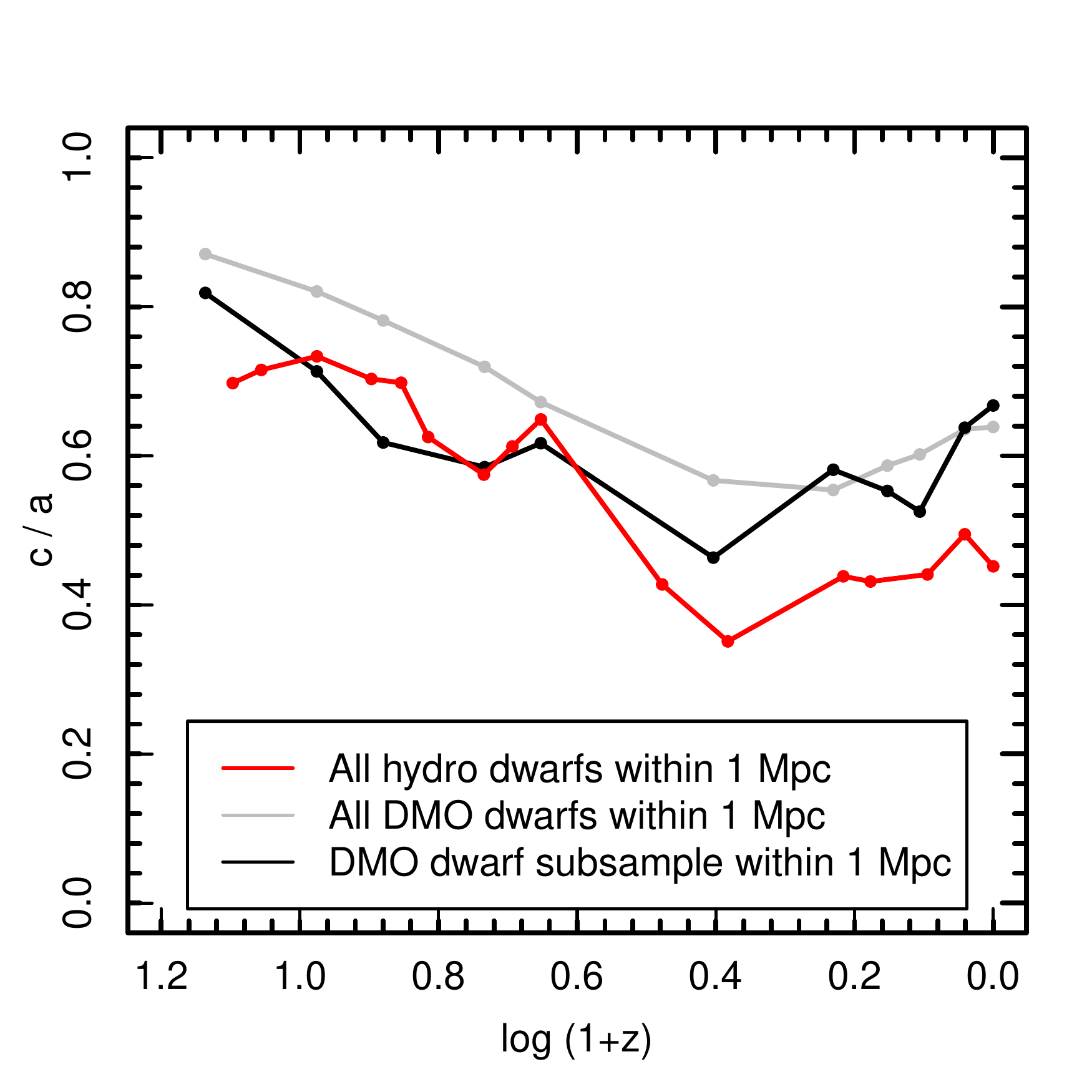}
\caption{A comparison of the spatial distribution of satellites, as indicated by the ``isotropy" $c/a$, at different redshift between the
Hydro and DMO cosmological simulations. 
We consider three satellite samples: the 11 most massive dwarfs within the virial radius (left panel), dwarfs within the virial 
radius ($R_{\rm vir}$) of the central galaxy (middle panel) , and dwarfs within 1 Mpc from the central galaxy (right panel). Note
in this Figure, the ``Hydro dwarfs" (in {\it red} in all three panels) refers to star-forming dwarfs  from the Hydro Simulation within
a given distance at different redshift (for convenient comparison, let $N_{zbar}$ be the number of these baryonic dwarfs at a given $z$),
the ``DMO dwarf subsample" (in {\it black} in the middle and right panels)  refers to a selective DMO dwarf sample which has the same number
as that of the Hydro dwarfs, the $N_{zbar}$ most massive ones from the DMO Simulation at the same redshift and within the same distance considered,
and ``All DMO dwarfs" (in {\it grey} in the middle and right panels) refers to all dwarfs formed from the DMO Simulation at the given redshift. All
distances are in comoving coordinates. 
}
 \label{fig:dmo_bar_11}
\end{figure*}

In order to understand the nature and the origin of the DoS, we analyze the satellites from two cosmological simulations of a MW-sized galaxy,
the Hydro Simulation with  comprehensive baryonic physics including star formation and feedback processes \citep{Marinacci2014}, and the DMO 
Simulation which is a pure N-body run \citep{Zhu2016}. We find that baryons can significantly affect the abundance and spatial distribution of
satellites \citep[see also][]{Zhu2016}. For example, within 1 Mpc, only $106$ luminous suhalos with star formation are found in the Hydro Simulation
and they are distributed anisotropically, in sharp contrast to the $\sim 21220$ subhalos which show isotropic distribution in the DMO Simulation.

Figure~\ref{fig:dmo_bar_11} shows the isotropy ratio $c/a$ of both simulations as a function of redshift for three samples: the 11 most massive dwarfs
within the virial radius (which have a 
similar mass range as the observed 11 ``classical'' satellites of the MW), dwarfs within the virial radius of the central galaxy, and dwarfs within 1
Mpc from the galaxy. 
These groups show three distinct trends in the evolution of $c/a$. When we select only the 11 most massive halos within the virial radius, the two
simulations show similar highly anisotropic distribution throughout time, and at $z=0$ the $c/a$ ratio is close to the observed value ($\sim 0.2$).
For dwarfs within $R_{\rm vir}$, the $c/a$ ratio generally increases as redshift approaches $z=0$ for all three samples i.e., the Hydro dwarfs, all
DMO dwarfs, and DMO dwarf subsample (massive DMO dwarfs with same sample size as the Hydro counterpart). 
This is mainly due to the rising abundance of 
dwarfs within the virial radius and phase mixing \citep{Henriksen1997}. The satellite infall near the center can be chaotic and even if some
satellites are accreted as a group from similar directions, as they move through the galactic potential, the neighboring satellites in phase-space
can become out of phase with time, resulting in a smooth 
phase-space distribution of satellites. This phase mixing is more effective for satellites closer to
the center \citep{Helmi2003}, which may explain the increased $c/a$ inside $R_{\rm vir}$.

On a galactic scale of 1~Mpc from the central galaxy, we find a remarkable difference between the Hydro and DMO simulations. At high redshift
($z \sim 10$) the satellite 
distributions are almost isotropic but over time the c/a ratio of both simulations declines,
although the decrease is much more significant in the Hydro simulation ($c/a \sim 0.4$ at $z=0$)
compared to the DMO one ($c/a \sim 0.64$ at $z=0$), even with the same sample size.

On a cosmic scale ($> 1$ Mpc), we find that both ratios ($b/a$ and $c/a$) continue to decrease, which suggest that the anisotropic dwarf
distribution may be part of the large scale filamentary structure. We discuss this in more detail in a companion paper (Maji et al., in prep).  
Our results suggest that on Mpc scales, the distribution of dwarfs around a central galaxy is anisotropic as part of the large-scale filamentary
structure. It has been suggested by many detailed 
DMO simulations that anisotropic satellite distribution can result from filamentary accretion of the satellites around the host galaxy
and the infall history can impact the final orientation of the satellites in the position-velocity space \citep{Aubert2004, Libeskind2005,
Libeskind2014, Lovell2011, Tempel2014, Buck2016}.

There are two factors responsible for the different satellite distributions between Hydro and DMO simulation: the difference in the 
satellite abundance and the effects of baryonic processes. Overall, the satellite abundance in DMO simulation is much higher than that in the
Hydro run, which in turn results in a more isotropic distribution, which is evident in Fig~\ref{fig:dmo_bar_11} (middle panel).
Furthermore, in the Hydro Simulation the dwarfs are subjected to additional baryonic processes, e.g. adiabatic contraction, tidal
disruption and reionization \citep{Zhu2016} that can significantly change the abundance,
star formation activity, infall time, and trajectory of the satellites. 
For very massive subhalos the effects are mild and the most massive halos in both 
simulations are essentially the same, resulting in very similar $c/a$ evolution for the 11 massive satellites. However,
for intermediate mass halos the tidal effects impacts the dynamics of the halos and even in similar mass range,
halos in DMO and Hydro simulation have different properties. Hence, in spite of having the same sample size, the halos within 1 Mpc shows a 
significantly different distribution for simulations with and without baryons. Similar results have also been suggested by recent 
studies  (e.g., \citealt{Ahmed2016, Zhu2016, Sawala2016, Zolotov2012}). Therefore, the inclusion of baryonic impacts may solve the 
discrepancy in the DoS anisotropy from previous
simulations \citep{Pawlowski2015methods, Sawala2016}.

\section{Conclusions}

In summary, we have performed a comprehensive reanalysis of the observed satellites of the MW using different plane 
identification methods and sample size. We have carried out Monte Carlo simulations to investigate the effects of 
sample size on the DoS properties, have calculated the future evolution of the 11 classical satellites in order to
test the stability of the current DoS, and have compared two cosmological simulations in order to understand the 
evolution of satellites and effects of baryons on the DoS properties. 
We find that the measured DoS properties strongly depends on the plane identification method and the sample size, 
and that a small sample size may artificially show high anisotropy and strong clustering. Furthermore, we find that
the DoS structure may be transient, and that baryonic processes play an important role in determining the distribution
of satellites. We conclude that the evidence for an ultra-thin, coherently-rotating DoS of the MW is not conclusive. 
Our findings suggest that the spatial distribution and kinematic properties of satellites may be  
determined by the assembly history and dynamical evolution of each individual galaxy system,
rather than being a universal DoS phenomenon.

\section{Acknowledgments}
YL acknowledges support from NSF grants AST-0965694, AST-1009867, 
AST-1412719 and MRI-1626251. We thank the anonymous referee, and Dr. Marcel 
Pawlowski and Dr. Yohan Dubois for thoughtful comments which have helped improve our 
manuscript. We acknowledge the Institute For CyberScience at The 
Pennsylvania State University
for providing computational resources and services that have contributed 
to the
research results reported in this paper. The Institute for Gravitation 
and the
Cosmos is supported by the Eberly College of Science and the Office of 
the Senior
Vice President for Research at the Pennsylvania State University.

\bibliography{dos_apjl.bib}

\end{document}